\documentclass[12pt]{article}
\def\rr{{{\bf r}}}
\def\rrp{{{\bf r^\prime}}}
\def\nup{{{n_\uparrow}}}
\def\ndo{{{n_\downarrow}}}

\begin{document}

\title{Stabilized jellium model and structural relaxation effects on the
fragmentation energies of ionized silver clusters }
\author{M. Payami}
\maketitle
\begin{center}
{\it Center for Theoretical Physics and Mathematics, Atomic Energy
Organization of Iran,\\ P.~O.~Box 11365-8486, Tehran, Iran}
\end{center}

\begin{abstract}
Using the stabilized jellium model in two schemes of `relaxed'
and `rigid', we have calculated the dissociation energies and the
fission barrier heights for the binary fragmentations of
singly-ionized and doubly-ionized Ag clusters. In the
calculations, we have assumed spherical geometries for the
clusters. Comparison of the fragmentation energies in the two
schemes show differences which are significant in some cases.
This result reveals the advantages of the relaxed SJM over the
rigid SJM in dynamical processes such as fragmentation. Comparing
the relaxed SJM results and axperimental data on fragmentation
energies, it is possible to predict the sizes of the clusters
just before their fragmentations.
\end{abstract}

\section{Introduction}

Metal clusters are one of the building blocks of the
nano-structured systems. The stability of these systems is of
great importance in nano-technology. One of the mechanisms which
can destroy the stability of metal clusters is charging. These
systems have been extensively studied\cite{brack93} in the context
of jellium model (JM). In the JM, the discrete ions are replaced
by a uniform positive charge background of density $n^B=3/4\pi
(r_s^B)^3$ in which $r_s^B$ is the bulk value of the Wigner-Seitz
(WS) radius of the valence electrons in the metal. A more
realistic version of the JM, the stabilized jellium model (SJM),
which was introduced \cite{pertran} by Perdew {\it et. al.} in
1990, has improved some drawbacks of the JM. The SJM calculations
for the fragmentation processes of metal clusters reported so far
\cite{Vieira98}, were based on the assumption that the density
parameter $r_s$ takes its bulk value $r_s^B$, which we call
`rigid' SJM. However, since the surface effects have large
contributions in the energetics and sizes of small clusters, a
more sophisticated use\cite{Perdew93} of the SJM is needed to
predict the correct energetics of the clusters in the study of
the fragmentation processes. This method, `relaxed' SJM, has
already been used to predict the equilibrium sizes and energies
of neutral as well as the charged\cite{PayamiJPC} metal clusters
(This method is also called self-compression method). In contrast
to the JM and the rigid SJM in which the $r_s$ value is borrowed
from the bulk system, in the relaxed SJM, the density parameter
of the jellium sphere assumes a value such that a cluster with a
given number of electrons and specific electronic configuration
achieves its equilibrium state. Comparing the relaxed SJM results
with the rigid SJM results on the one hand, and the experiment on
the other hand, provide information on the possible structural
relaxations of the clusters in the fragmentation processes.

In this work, we have studied the binary decay processes of
positively charged silver clusters Ag$_N^Z$ ($Z$=1, 2) containing
up to 100 atoms, in all possible channels, using the relaxed and
the rigid SJMs. The possible decay channels for singly-ionized  Ag
clusters are

\begin{equation}
{\rm Ag}_N^{1+}\to {\rm Ag}_{N-p}^{1+} + {\rm Ag}_p^0,
\;\;\;\;\;\;\;\;p=1,2,\cdots,N-2. \label{eq1}
\end{equation}

For doubly charged clusters, the decays can proceed via two
different processes. The first one is the evaporation process

\begin{equation}
{\rm Ag}_N^{2+}\to {\rm Ag}_{N-p}^{2+} + {\rm Ag}_p^0,
\;\;\;\;\;\;\;\;p=1,2,\cdots,N-3, \label{eq2}
\end{equation}
in which one of the products is neutral; and the second one is
fission into two charged products

\begin{equation}
{\rm Ag}_N^{2+}\to {\rm Ag}_{N-p}^{1+} + {\rm Ag}_p^{1+},
\;\;\;\;\;\;\;\;p=2,3,\cdots,[N/2]. \label{eq3}
\end{equation}

In the evaporation processes, the negativity of the difference
between total energies before and after fragmentation in a
specific channel,

\begin{equation}
 D^Z(N,p)=E^Z(N-p) + E^0(p) - E^Z(N), \label{eq4}
\end{equation}
is sufficient for the decay in that channel to take place. In the
above equation, $E^Z(N)$ and $E^0(N)$ are the total energies of
$Z$-ply ionized and neutral $N$-atom clusters, respectively.
However, in fission processes a negative value for the difference
energy is not a sufficient condition for the decay of the parent
cluster. This is because the competition between the short-range
surface tension and the long-range repulsive Coulomb force may
give rise to a fission barrier. The height of the fission
barriers are calculated using the two-spheres
approximation\cite{Naher}. The situation in a fission process is
shown in Fig. \ref{fig1}. In figure \ref{fig1}, the fission of a
$Z$-ply charged $N$-atom cluster into two clusters of respective
sizes $N_1$, $N_2=N-N_1$ and respective charges $Z_1$,
$Z_2=Z-Z_1$ is schematically shown. $Q_f$ is the energy release,
$B_c$ is the fusion barrier which is the maximum energy of the
Coulomb interaction of two positively-charged conducting spheres,
taking their polarizabilities into account. $B_f$ is the fission
barrier height which is defined as

\begin{equation}
B_f=-Q_f+B_c. \label{eq5}
\end{equation}

The coulomb interaction energy, $E_c$, as a function of their
separations, $d$, for two charged metal spheres can be
numerically calculated using the classical method of image
charges \cite{Naher}. Our calculations show that the maximum of
the interaction energy, $B_c$, is achieved for a separation
$d_0\ge R_1+R_2$. The equality applies for equal cluster radii and
charges. The electrostatic interaction energies of pairs of
equally charged ($Z_1=Z_2=1$) clusters as functions of the pair
separation are shown in Fig. \ref{fig2} for different pair sizes.
The radii of the clusters in this figure are calculated from
$R=N^{1/3}r_s^B$. In Fig. \ref{fig3}, the values of the Coulomb
barrier, $B_c$ are plotted for different pair sizes but equal
charges $Z_1=Z_2=1$ in the rigid SJM. As is seen, when both of the
cluster sizes are small, the barrier is higher than the case when
at least one of them is larger.

The most favored decay channel in evaporation processes is
defined as the channel for which the dissociation energy attains
its minimum value

\begin{equation}
D^Z(N,p^*)=\min\left\{D^Z(N,p)\right\},\label{eq6}
\end{equation}
and the most favored decay channel in fission processes is
defined as the channel for which the fission-barrier height
attains its minimum value

\begin{equation}
B_f(N,p^*)=\min\left\{B_f(N,p)\right\}.\label{eq7}
\end{equation}

\section{Total energies of clusters}

The total energy of an $N$-electron $Z$-ply charged cluster is
obtained by solution of the self-consistent Kohn-Sham (KS)
equations \cite{KohnSham} in the density functional\cite{Kohn64}
theory (DFT) with local spin density approximation (LSDA) for the
exchange-correlation energy functional. The SJM energy for a
cluster in the LSDA is given by\cite{PayamiJPC}

\begin{eqnarray}
E_{\rm SJM}\left[\nup,\ndo,n_+\right]&=& E_{\rm
JM}\left[\nup,\ndo,n_+\right]+\left(\varepsilon_M(r_s)+\bar
w_R(r_s,r_c^B)\right)\int d\rr\;n_+(\rr) \nonumber \\
  &&+\langle\delta v\rangle_{\rm WS}(r_s,r_c^B)\int
d\rr\;\Theta(\rr)\left[n(\rr)-n_+( \rr)\right], \label{eq8}
\end{eqnarray}
where
\begin{eqnarray}
E_{\rm
JM}\left[\nup,\ndo,n_+\right]&=&T_s\left[\nup,\ndo\right]+E_{xc}\left[\nup,\ndo
\right] \nonumber\\ &&+\frac{1}{2}\int
d\rr\;\phi\left([n,n_+];\rr\right)\left[n(\rr)-n_+(\rr)\right]
\label{eq9}
\end{eqnarray}
and
\begin{equation}
\phi\left([n,n_+];\rr\right)=\int
d\rrp\;\frac{\left[n(\rrp)-n_+(\rrp)\right]}{\left|\rr-\rrp\right|}.
\label{eq10}
\end{equation}
Here, $n=n_\uparrow+n_\downarrow$ which satisfies $\int d\rr
n(\rr)=N-Z$, and $n_+$ is the jellium density which satisfies
$\int d\rr n_+(\rr)=N$. $\Theta(\rr)$ takes the value of unity
inside the jellium background and zero, outside. The first and
second terms in the right hand side of Eq. (\ref{eq9}) are the
non-interacting kinetic energy and the exchange-correlation
energy, and the last term is the Coulomb interaction energy of
the system. $\varepsilon_{\rm M}$ is the average Madelung energy.
All equations throughout this paper are expressed in atomic units
($\hbar=e^2=m=1$, the units of length and energy are bohr and
hartree, respectively). The quantity $\langle\delta v\rangle_{\rm
WS}$ is the average of the difference potential over the
Wigner-Seitz cell and the difference potential, $\delta v$, is
defined as the difference between the pseudo-potential of a
lattice of ions and the electrostatic potential of the jellium
positive background \cite{pertran}.

The SJM ground-state energy [Eq. (\ref{eq8})] for a cluster with
$N$ electrons is a function of $N$, $r_s$, and $r_c^B$. In the
rigid SJM, $r_s$ takes the bulk value $r_s^B$, whereas in the
relaxed SJM, it takes the equilibrium value, $\bar r_s(N)$, which
for a cluster in the ground state electronic configuration, is
obtained by the solution of the equation

\begin{equation}\label{eq11}
\left.\frac{\partial}{\partial r_s}E_{\rm
SJM}(N,r_s,r_c^B)\right|_{r_s=\bar r_s(N)},
\end{equation}
where, the derivative is taken at fixed values of $N$ and $r_c^B$.
The total energies of the cluster in the relaxed SJM and the rigid
SJM are given by $\bar E_{\rm SJM}=E_{\rm SJM}(N,\bar
r_s(N),r_c^B)$ and $E_{\rm SJM}(N,r_s^B,r_c^B)$, respectively.

\section{Results and discussion}

To obtain the relaxed SJM properties, we have performed an
extensive self-consistent solutions of the KS equations along
with Eqs. (\ref{eq8})-(\ref{eq11}), and have calculated the
equilibrium $r_s$ and the energy values of Ag$_N^Z$ clusters
($Z=0,1,2$) for different cluster sizes ($N\le 100$). To show the
main differences in the equilibrium $r_s$ values of these
clusters, which are appreciable for small clusters, we have
plotted in Fig. \ref{fig4} the corresponding values only up to
$N=34$. As is obviously seen in the figure, the neutral and
singly-ionized clusters are self-compressed for all values of
$N$.  However, for doubly-ionized clusters, the values cross the
bulk border (i.e., $r_s^B=3.02$) at $N=7$.

Figure \ref{fig5} shows the relaxed SJM energies per atom in
electron volts for neutral, singly-ionized, and doubly-ionized
silver clusters with $N\le 34$. For comparison, we have also
plotted the bulk value ($\varepsilon=-7.89\;eV$) by a dashed
line. Using these values of equilibrium energies and sizes, the
relaxed SJM fragmentation energies are calculated by the Eqs.
(\ref{eq5})-(\ref{eq7}).

In Fig. \ref{fig6}, we have plotted the dissociation energies of
the most favored evaporation channels of the singly ionized
clusters. We have shown the most favored value of $p$ by $p^*$.
The solid small square symbols show the most favored values $p^*$
on the right vertical axis whereas, the corresponding
dissociation energies, $D^{1+}(N,p^*)$, are shown on the left
vertical axis by large open squares. As is seen in the figure,
there exist some maxima and minima. The maxima of the
$D^Z(N,p^*)$ correspond to the closed-shell Ag$_N^+$ clusters
with $N$=3, 9, 19, 35, 59,$\ldots$. These clusters have high
stabilities compared to their neighboring sizes. On the other
hand, the minima correspond to the sizes which decay into two
closed-shell clusters (for example, Ag$_{11}^+\rightarrow$Ag$_9^+
+$Ag$_2$). A negative value for the dissociation energy implies
that the cluster is unstable against the spontaneous decay.

In Fig. \ref{fig7}(a), we have shown the most favored products
Ag$_{p^*}^0$ and the dissociation energies $D^{2+}(N,p^*)$ for the
decay of {\rm Ag}$_N^{2+}$ via evaporation channel. It is seen
that, here, the mean dissociation energy is higher than that in
the evaporation of singly ionized clusters. That is, here, the
number of clusters stable against the spontaneous evaporation is
larger than that in the singly ionized case. However, evaporation
is not the only decay mechanism for multiply charged clusters, and
they can also decay via fission processes in which both fragments
are charged.

Figure \ref{fig7}(b) shows the barrier heights $B_f(N,p^*)$ for
the most favored fission channels ${\rm Ag}_N^{2+}\to {\rm
Ag}_{N-p^*}^{+} + {\rm Ag}_{p^*}^{+}$. Here, in the relaxed SJM,
the Coulomb barriers are calculated using the equilibrium sizes
of each $Z$-ply charged fragments i.e., $\bar R^Z(N)=N^{1/3}\bar
r^Z_s(N)$. As is seen, for small clusters, the majority have
negative barrier heights. That is, most of them are unstable
against spontaneous fission. However, as $N$ increases, the
number of clusters with negative barrier heights decreases and
beyond a certain size range, all the barrier heights become
positive. As before, the fragment sizes $p^*$ are shown in the
left vertical axis.

As mentioned before, a doubly charged cluster can decay both via
evaporation and fission. To estimate at what sizes which decay
mechanism is dominant, we have compared the dissociation energies
and the fission barrier heights in Fig. \ref{fig7}(c). At small
sizes ($N<21$), the fission process dominates because, the
barrier heights for the fission are lower than the dissociation
energies for the evaporation. However, the competition between
the evaporation and fission starts at $N$=21. This competition
continues with some fluctuations until the evaporation dominates
completely. Ag clusters of sizes $\sim 1\;nm$ and $\sim 2\;nm$
contain $\sim 40$ and $\sim 300$ atoms, respectively. Knowing the
fact that, in doubly ionized clusters, the induced evaporation is
most favored for $N>21$, we conclude that the ionized clusters of
nano-meter size undergo the fragmentation mostly via the
evaporation processes.

The rigid SJM fragmentation energies have been calculated by the
self-consistent solutions of the KS equations for the system with
the positive background density parameter of $r_s^B$. In Figs.
\ref{fig8}(a)-(c), the corresponding fragmentation energies of
the relaxed and the rigid SJM are compared. As is seen, in some
cases the energy difference is as large as 0.5 eV, indicating the
significant effect of relaxation.

Improvements over our relaxed SJM results can be obtained by
assuming that, at the instant the fragmentation takes place, the
the parent and the products are not at their individual
equilibrium state but that, the $r_s$ values of the products are
equal to that of the parent at that instant. Thus, in the
equations used for the dissociation energies and the fission
barrier heights this fact must be taken into account. The $r_s$
value of the parent cluster at the fragmentation instant (i.e.,
at the instant the two pieces are created) is something which is
between that of the neutral cluster with the same $N$ and that of
the relaxed $Z$-ply charged of the same $N$. That is,
$r_s^0(N)\le r_s \le r_s^{2+}(N)$ for the fission of a doubly
charged $N$-atom cluster. The reason for this fact lies in the
probabilistic nature of the energy absorption in the ionization
process of the neutral cluster as well as the energy absorption
in the fragmentation process of the $Z$-ply charged parent
cluster. If the fragmentation energy is absorbed as soon as the
electrons are detached from the neutral cluster, then the parent
charged cluster does not have enough time for structural
relaxation due to its charging. Therefore, the appropriate $r_s$
value would be that of the neutral cluster. On the other hand, if
the absorption of the fragmentation energy takes place with
enough delay relative to the ionization instant, then the parent
charged cluster has a chance to arrive at its relaxed $r_s$
value. These arguments lead us to conclude that the experimental
fragmentation energies are highly dependent on the details of the
experimental setup. This freedom in taking the $r_s$ values of
the clusters in a dynamical process such as fragmentation shows
the advantages of the relaxed SJM over the rigid one. However, in
cases where the relaxed and the rigid $r_s$ values are more or
less the same, one would expect that applying the time-saving
rigid SJM would lead to the same estimates of the fragmentation
energies.

\section{Summary and conclusions}

In this work, we have calculated the dissociation energies and
the fission barrier heights of the fragmentations of singly and
doubly ionized silver clusters in the two schemes of relaxed and
rigid SJMs for different cluster sizes, $N\le 100$. In our
calculations, we have assumed spherical geometries for the
clusters. In the rigid SJM, the density of the positive charge
background is taken to be that of the bulk system, $n^B$.
However, in the relaxed SJM, that density is obtained in a
self-consistent way which corresponds to the equilibrium state of
the cluster. The fission barrier heights are calculated using the
two-spheres approximation and for the Coulomb interaction of the
two metallic spheres we have employed the method of image charges
numerically. The differences in the results of the two schemes
show the advantages of the relaxed SJM over the rigid one in that,
one is able to take into account the structural relaxations in
dynamical processes such as fragmentations.

\newpage

\begin{figure}
\caption{Fission barrier in the two-spheres approximation. The
parent $N$-atom $Z$-ply charged cluster decays into two clusters
of sizes $N_1$ and $N-N_1$, with charges $Z_1$ and $Z_2$,
respectively. } \label{fig1}
\end{figure}

\begin{figure}
\caption{Coulomb energy in electron volts of two singly charged
metallic spheres as a function of their separation
distance.}\label{fig2}
\end{figure}

\begin{figure}
\caption{Coulomb barrier heights in electron volts for two singly
charged metallic spheres with different sizes.}\label{fig3}
\end{figure}

\begin{figure}
\caption{Equilibrium $r_s$ values in atomic units for neutral,
singly ionized, and doubly ionized silver clusters. The
horizontal dashed-line shows the bulk value for silver, 3.02.
}\label{fig4}
\end{figure}

\begin{figure}
\caption{Equilibrium energies in electron volts of neutral,
singly ionized, and doubly ionized silver clusters. The
horizontal dashed-line shows the bulk value for silver, -7.89
$eV$. }\label{fig5}
\end{figure}

\begin{figure}
\caption{Dissociation energies, in electron volts, of the most
favored decay channels of singly ionized clusters, in relaxed SJM,
are shown with respect to the left vertical axis. The right
vertical axis shows the sizes of the fragments in the most favored
channels.} \label{fig6}
\end{figure}

\begin{figure}
\caption{Relaxed SJM results for doubly ionized silver clusters.
a)- the same quantities as in Fig.\ref{fig6} for doubly ionized
clusters; b)- the fission barrier height in electron volts for
the most favored fission channel as a function of the cluster
size. The right vertical axis shows the most favored product
sizes of the neutral clusters; c)- comparison of the decay
energies of via evaporation and fission mechanisms. Competition
starts at $N=21$ } \label{fig7}
\end{figure}

\begin{figure}
\caption{Comparison of the most favored channel fragmentation
energies, in electron volts, for the two schemes of the relaxed
and rigid SJMs. a)- dissociation energies of singly ionized
clusters; b)- dissociation energies of doubly ionized clusters;
c)- fission barrier heights for the fission of doubly ionized
clusters into two singly ionized clusters.}\label{fig8}
\end{figure}

\end{document}